\documentclass[aps,prl,superscriptaddress,showpacs,twocolumn,longbibliography]{revtex4-2}
\usepackage{amsmath, bm}
\usepackage{amssymb}
\usepackage{graphicx}
\usepackage{hyperref}
\usepackage{color}
\usepackage{ulem}
\usepackage{hhline}
\usepackage{subeqnarray}
\usepackage{array}
\usepackage{subfigure}

\graphicspath{{./FIGURES/}}

\hypersetup{breaklinks=true,colorlinks=true}

\begin{document}
\title{Chiral order emergence driven by quenched disorder}

\author{Coraline Letouz\'e}
\affiliation{Sorbonne Universit\'e, CNRS, Laboratoire de Physique Th\'eorique de la Mati\`ere Condens\'ee (LPTMC), F-75005 Paris, France}

\author{Pascal Viot}
\email{pascal.viot@sorbonne-universite.fr}
\affiliation{Sorbonne Universit\'e, CNRS, Laboratoire de Physique Th\'eorique de la Mati\`ere Condens\'ee (LPTMC), F-75005 Paris, France}

\author{Laura Messio}
\email{laura.messio@sorbonne-universite.fr}
\affiliation{Sorbonne Universit\'e, CNRS, Laboratoire de Physique Th\'eorique de la Mati\`ere Condens\'ee (LPTMC), F-75005 Paris, France}

\begin{abstract}
Quenched disorder can destroy magnetic order, for example when a random field is applied in a 2-dimensional Ising model. 
Even when an order exists in the presence of quenched disorder, it is usually only the survival of the order of the clean model.
We present here a surprising phenomenon where an order emerges, driven by quenched disorder. 
This order has nothing in common with the order present in the clean model. 
This type of \textit{order by disorder} differs from the usual thermal or quantum one. 
The classical $J_1-J_3$ Heisenberg model on the kagome lattice is studied by parallel tempering Monte Carlo simulations, with site dilution.
After analyzing the effect of a few vacancies on the ground state, favoring non-coplanar configurations, we show the emergence of a low-temperature chiral phase and the progressive destruction of the collinear $q=4$ Potts order, the only order present in the absence of vacancies. 
\end{abstract}

\date{\today}

\maketitle

In two-dimensional frustrated magnets with continuous spin degrees of freedom, non-collinear or non-coplanar spin orders may break discrete symmetries, such as time-reversal or lattice symmetries. 
Then the Mermin-Wagner theorem allows for finite-temperature phase transitions,  resulting in chiral orders, stripe patterns, or valence bond crystals \cite{Hasenbusch2006,MessioDomenge}. 
The symmetries of the low-temperature phase can usually be inferred from the structure of the ground-state manifold. When these states are not all related by the Hamiltonian's symmetries, one speaks of accidental degeneracy, as observed in classical spin liquids \cite{PhysRevB.110.L020402}. 
At finite temperatures, entropic contributions favor ground states with a high density of low-energy excitations, a mechanism known as order by thermal disorder (ObTD)\cite{VillainJ.1980}. 
Similarly, quantum fluctuations (finite spin value) give rise to order by quantum disorder (ObQD). The selected states are typically the \textit{most-collinear} spin configurations\cite{Henley1987, Henley1989, Maryasin2013}. 
Several paradigmatic models illustrate the thermal or quantum ObD, among them 
the $J_1-J_2$ $XY$\cite{Henley1989} or Heisenberg\cite{Chandra1990,Weber2003} model on the square lattice where fluctuations favor collinear spin configurations, generating an Ising-like phase transition at finite temperature, 
the kagome antiferromagnet, where coplanar spin configurations are selected\cite{Chalker1992, Moessner1998, Zhitomirsky2008} 
and the $XY$ pyrochlore lattice\cite{PhysRevB.68.020401}, where a complex non-coplanar order is selected.

Quenched (structural) disorder, such as bond or site disorder, is generally expected to suppress or destabilize any kind of order. 
Phase transitions in spin models with quenched disorder typically fall into one of these two classes, according to the symmetry of the order parameter and the nature of the disorder \cite{Vojta2019}:
(i) Random-field disorder, which explicitly breaks the symmetry between ordered states, typically precluding long-range order in dimensions $d \leq 2$ for a discrete order, $d\leq4$ for a continuous order, by the Imry-Ma argument \cite{PhysRevLett.35.1399}; and
(ii) Random-$T_c$ disorder, which preserves the symmetry of the ordered phase but induces spatial fluctuations in the critical temperature\cite{Harris1974a}. 
For example, in the 2d Ising model, random-fields suppress the transition\cite{PhysRevLett.35.1399}, while site dilution (random-$T_c$ disorder) just reduces it, lowering the critical temperature\cite{Dotsenko01011983, SELKE1998388}. 
On the $J_1$–$J_2$ Ising model on the square lattice, site dilution affects the low-temperature phases in distinct ways according to the order parameter, preserving ferromagnetic order (random-$T_c$ disorder) while destroying the stripe phase (random-field disorder), by locally breaking the $C_4$ lattice symmetry \cite{PhysRevB.98.024206,Vojta2019,PhysRevB.105.024201} and creating effective random fields.

However, when the order emerges from subtle phenomena such as ObTD or ObQD, quenched disorder not only destabilize it, but may induce the emergence of new orders. By opposition to ObTD or ObQD, quenched disorder is known to favor the \textit{least-collinear} spin configurations\cite{Henley1987, Henley1989, PhysRevLett.70.3812, Maryasin2013, Andreanov2015,  PhysRevB.90.094412, PhysRevLett.119.047204, Andrade2018}.

\begin{figure}[t]
\includegraphics[width=.45\textwidth]{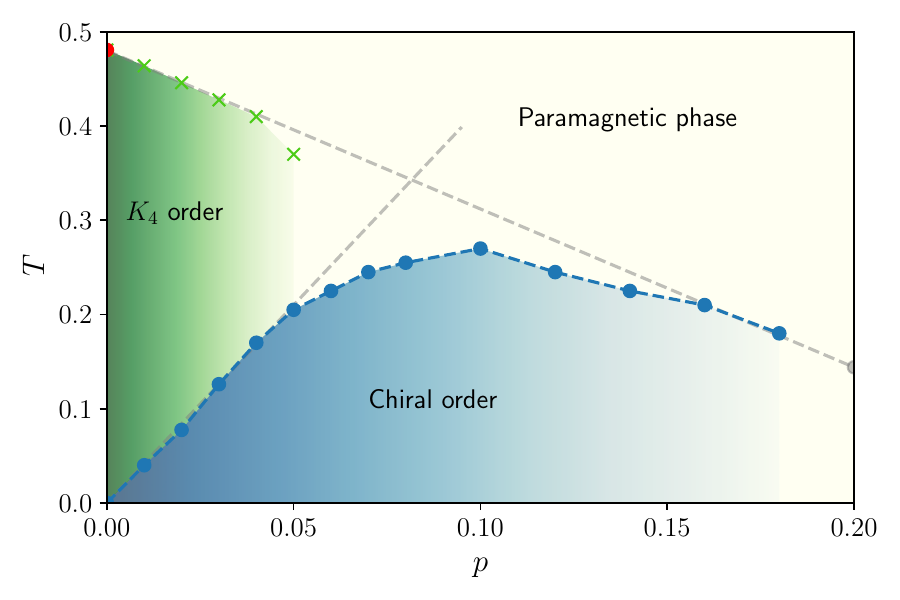}
\caption{Phase diagram of the $J_1J_3$HK model with $J_1 = -1$ and $J_3 = 1$.
$p$ denotes the dilution rate and $T$ the temperature.
The blue region indicates a phase with non-zero scalar chirality $\chi\neq 0$, while the green region corresponds to non-zero $K_4$ order parameter $\sigma\neq 0$.
Blue and red dots indicate phase transitions, while green crosses correspond to a crossover. }
\label{fig:PhaseDiag}
\end{figure}

In this article, we present an $O(3)$ symmetric model where quenched disorder induces the emergence of a chiral phase: the $J_1-J_3$ Heisenberg model on the kagome lattice ($J_1J_3$HK model). 
This phase, spontaneously breaking the time-reversal symmetry at low temperatures, gives rise to a finite-temperature phase transition as soon as dilution is present, see Fig.~\ref{fig:PhaseDiag}. 
Remarkably, this phenomenon appears to be robust and potentially general in several Heisenberg models. 
This discovery is particularly noteworthy, as the resulting chiral phase is fundamentally distinct from the discrete 4-state Potts order (denoted $K_4$ in the following, as the Klein four-group) stabilized via ObTD\cite{Grison2020}. 
Crucially, it does not merely arise from a symmetry reduction that selectively favors a subset of ground states \cite{PhysRevB.105.024201}, but represents a qualitatively different form of order induced by the presence of quenched disorder.
Although, due to vacancies, the apparition of a chirality has already been reported in a $XY$ model in the seminal work of Henley\cite{Henley1989}, as well as the apparition of new phases in Heisenberg models with a magnetic field\cite{Maryasin2013}, this is to our knowledge the first mention of the emergence of a discrete chirality in a model with O(3) spin-rotation symmetry. 

\begin{figure}
	\centering
	\includegraphics[height=.23\textwidth]{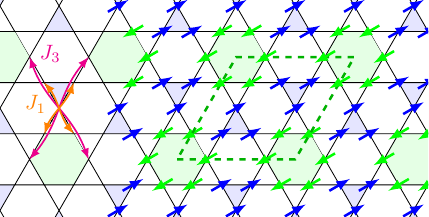}
	\caption{Left: First ($J_1$) and third-neighbor ($J_3$) interactions in the $J_1J_3$HK model.
		Right: spin orientations of the collinear $K_4$ order, characterized by a four-site unit cell (dashed outline). 
    Sites with the same spin orientation form hexagons (light green) and triangle (light blue). }
	\label{fig:neighbor_kagome}
\end{figure}

The $J_1J_3$HK model features ferromagnetic nearest-neighbor Heisenberg interactions ($J_1$) and antiferromagnetic third-neighbor couplings ($J_3$), as illustrated in Fig.~\ref{fig:neighbor_kagome}.
This model with no $J_2$ coupling yet large $J_3$ may appear unphysical and of purely theoretical interest.
In fact, this model is the classical approximation of the $S=1/2$ $J_1J_3$HK model describing the Ba-Vesignieite compound BaCu$_3$V$_2$O$_8$(OH)$_2$\cite{JPSJ.78.033701,PhysRevB.83.180416,PhysRevB.83.180407,c2jm32250a,Ishikawa2017} where the $J_3$ coupling dominates\cite{PhysRevLett.121.107203}. 

In the presence of site dilution, the Hamiltonian reads:
\begin{equation}
	\label{eq:HamJ1J3}
	\mathcal{H}=
	J_1\sum_{\langle i,j\rangle}\eta_i\eta_j \mathbf S_i \cdot \mathbf S_j
	+J_3\sum_{\langle i,j\rangle_3} \eta_i\eta_j\mathbf S_i \cdot \mathbf S_j,
\end{equation}
where $\mathbf{S}_i$ are classical 3-component unit vectors, and $\eta_i = 1$ or $0$ indicates the presence or absence of a magnetic spin at site $i$, respectively. The $\eta_i$ are uncorrelated random variables representing quenched disorder. 
The dilution rate is defined as $p = 1 - \sum_i \eta_i/N$, where $N$ is the total number of sites.

\begin{figure}
	\centering
	\includegraphics[height=.23\textwidth]{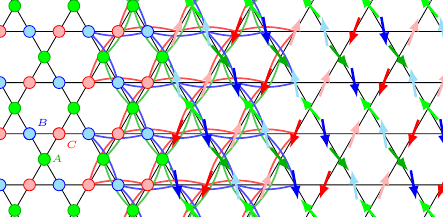}
	\caption{Spin structure in the 3subAF phase on the kagome lattice.
		Circles denote the three sublattices $A$, $B$, and $C$, connected via third-neighbor $J_3$ couplings (colored links).
		Each sublattice independently exhibits Néel order with two opposing spin orientations (light/dark arrows, of the color of the sublattice).
		The collinear limit of this structure corresponds to the $K_4$ order shown in Fig.~\ref{fig:neighbor_kagome}.}
	\label{fig:Kag_6sublattices}
\end{figure}

The $J_1J_3$HK model without site dilution exhibits a rich zero-temperature phase diagram \cite{Grison2020,Li2022}. 
For $J_1=0$, the kagome lattice decouples into three independent (squeezed) and equivalent square sublattices (in different colors on the left of Fig.~\ref{fig:Kag_6sublattices}), each displaying ferromagnetic ($J_3<0$) or antiferromagnetic ($J_3>0$) ordering. 
This results in either a 3-sublattice or 6-sublattice configuration, depending on the sign of $J_3$ (see right of Fig.~\ref{fig:Kag_6sublattices}). 
Beyond global $O(3)$ spin rotation symmetry, the model hosts an accidental degeneracy which is not lifted by small $J_1$ values for $J_3 > 0$. 
Then the system resides in the so-called 3subAF phase. 
Fixing $J_3 = 1$, this phase remains stable over a wide interval: $ -1.24 \lesssim J_1 < 1$ \cite{Grison2020}.
Each configuration in the 3subAF phase is fully characterized by 3 spin orientations $(\mathbf S_A, \mathbf S_B, \mathbf S_C)$ on a reference triangle $ABC$ of the lattice, as it allows to propagate these orientations and their opposites on the whole lattice (Fig.~\ref{fig:Kag_6sublattices}). 
Global spin rotation are symmetries of the model, thus ground states related by such transformations are considered as equivalent (this continuous symmetry is only broken at $T=0$ according to the Mermin Wagner theorem). 
The equivalence class of a ground state is is thus specified by a couple $(\bm\sigma,\chi)$ of a three dimensional real vector $\bm\sigma$ and a chirality $\chi=\pm1$ defined on a triangle $ABC$ as:
\begin{equation}
	\label{eq:orderParameters}
	\bm\sigma_{ABC} 
	= \begin{pmatrix}\mathbf S_B\cdot \mathbf S_C\\ \mathbf S_C\cdot \mathbf S_A\\ \mathbf S_A\cdot \mathbf S_B\end{pmatrix},\quad
	\chi_{ABC}= \mathbf S_A\cdot (\mathbf S_B\land \mathbf S_C). 	
\end{equation}

Among these classes, 4 have collinear spins $\pm\mathbf S_A=\pm\mathbf S_B=\pm\mathbf S_C$ resulting in 4 possible values for $\bf\sigma$ and $\chi=0$ (represented on Fig.~\ref{fig:neighbor_kagome} right) and 2 others have orthogonal spins pointing to the vertices of an octahedron ($\bf\sigma=0$ and $\chi=\pm1$). 
The configurations with orthogonal spins are called octahedral. 
These two chiral spin states (among all other chiral ones) have the particularity to not break any lattice symmetry\cite{PhysRevB.83.184401}.

At low temperature and for $p=0$, the 4 collinear classes of the 3subAF states are favored through ObTD\cite{Grison2020}. 
It induces a discrete symmetry breaking related to the choice of one of the four possible classes (emergent $q = 4$ Potts, or $K_4$ order parameter). 
The order parameter for this order is denoted $\Sigma$ and is defined from $\sigma$ (Eq.~\ref{eq:orderParameters}) similarly to the alternated magnetization of a Néel order\cite{Grison2020}, while the chiral order parameter is $\chi$: the absolute value of the average of $\chi_{ABC}$ over the lattice. 
Finite-size analysis have shown that the critical exponents are compatible with the $4$ state Potts model universality class\cite{Grison2020}.
Notably, there is a strong similitude with the $J_1-J_2$ Heisenberg model on the square lattice ($J_1J_2$HS), where the large-$J_2$ limit consists in 2 decoupled square lattices and only 2 collinear orders (distinct up to global spin rotations), leading to an Ising-like order. 

What are now the effect of a non-zero (small) $p$ in the $J_1J_3$HK model ? And first of all, does the $K_4$ order survive such quenched disorder ? First, we discuss the emergence of random fields destroying the $K_4$ order, then the apparition of chiral configurations and finally, we present the numerical simulations evidencing the chiral transition and the evolution of the $K_4$ phase transition.
In the $J_1J_3$HK model, although the disorder at hand is not a random field \textit{per se} but site disorder, it still generates an effective random field, because both the $K_4$ order and the site vacancies break the translational lattice symmetries\cite{Vojta2019}. 
To illustrate this, we evaluate the effect of a small number of vacancies on the collinear spin configuration of Fig.~\ref{fig:neighbor_kagome} in the approximation where the state does not reorder to minimize its energy. 
It happens that a hole on a hexagon of green spins has the same energetic effect as a hole on a triangle of blue spins. 
However, as soon as two neighboring sites are empty, the energy difference with respect to the fully occupied state depends on the hole position: $8J_3-J_1$ for two sites of different colors versus $8J_3+J_1$ for two identical-color sites. 
Site disorder thus affects the 4 translationally equivalent collinear orders differently, similar to what a random field does on otherwise equivalent magnetizations. 
According to the Imry-Ma argument, the $K_4$ order is thus expected to be killed by an infinitesimal $p>0$, as we are at the critical lower dimension $d=2$ for a discrete order\cite{PhysRevLett.35.1399}.
 
\begin{figure}
	\centering
	\includegraphics[width=0.49\textwidth]{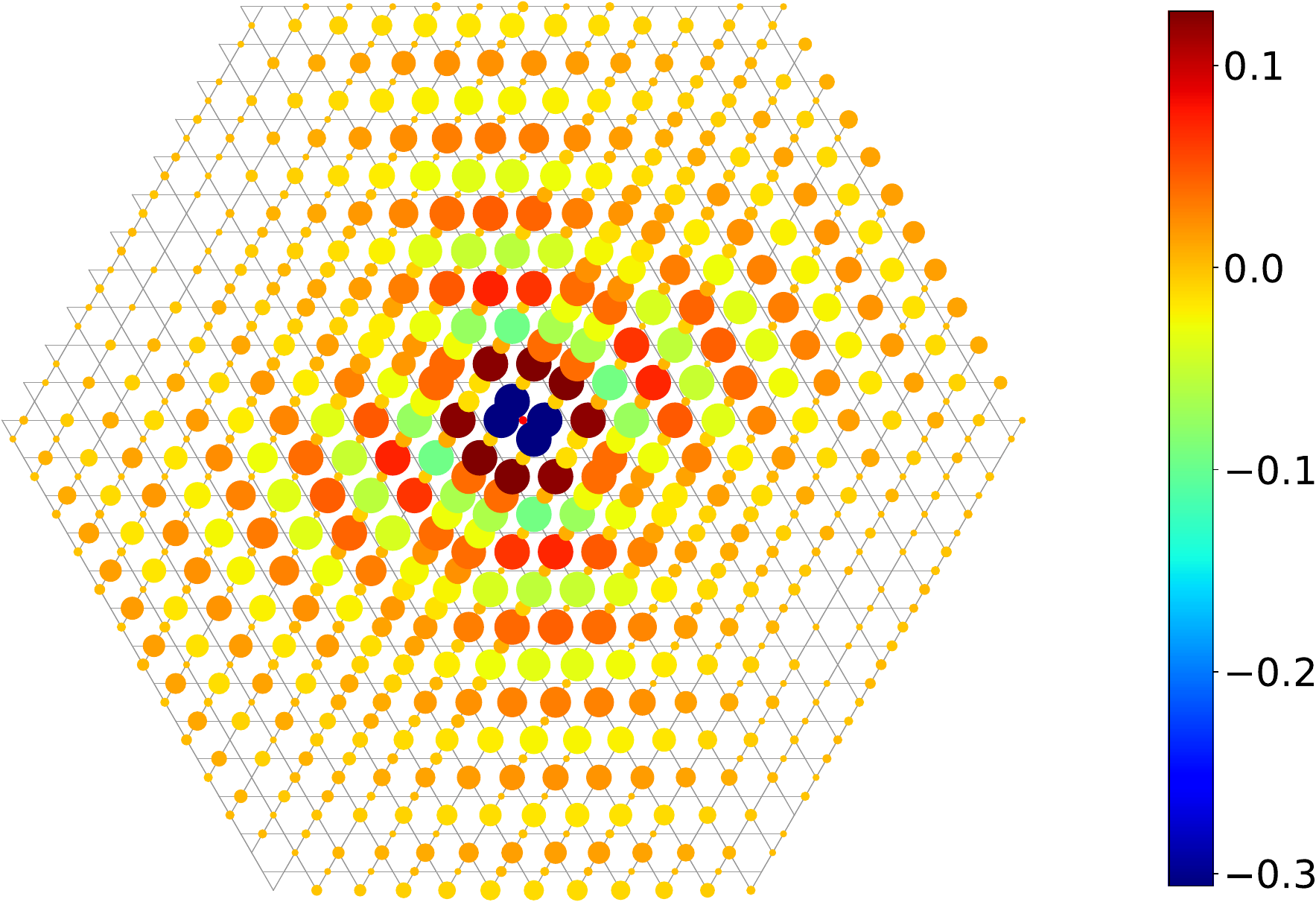}
	\caption{Ground state spin-spin correlation $\mathbf S_i\cdot \mathbf S_0$ between sites of the $A$ and $C$ sublattices and the central site $0$ belonging to the $B$ sublattice, for a periodic $24\times24\times3$ lattice. 
	The exchanges of the $J_1$ and $J_3$ links emanating from site $0$ tend to zero. 
	Correlations between $B$ sites are not shown (they are strongly antiferromagnetic and near $\pm1$), and the diameter of circles saturate for absolute values above 0.04}
	\label{fig:Correlations_1imp}
\end{figure}

At very low temperature, the effective fields not only kill the $K_4$ order, but allows the emergence of a new discrete order parameter, related to the chirality sign. 
Unlike $K_4$, chirality does not break any lattice symmetry and does not suffer from any random field effect, fully allowing a phase transition in the presence of site disorder. 
To illustrate this, we begin by considering a single vacancy on a $B$-site of the kagome lattice (Fig.~\ref{fig:Correlations_1imp}). 
For $J_1=0$, the presence or absence of this spin has strictly no effect on the ground state manifold, but switching on $J_1$ lifts the degeneracy. 
The $A$ and $C$ spins of the four first neighbors of the impurity experience an apparent magnetic field of strength $J_1$, in the direction opposite to the missing spin\cite{Henley2001}. 
In a N\'eel antiferromagnet, parallel and perpendicular magnetic susceptibilities differ: $\mathcal S_\parallel=0$ while $\mathcal S_\perp>0$: the spins tend to orient perpendicularly to the magnetic field and to minimize their energy by a canting that decreases when moving away from the impurity. 
As a consequence, the $A$ and $C$ Néel ordered sub-lattices orient quasi-perpendicularly to the $B$ spins, with angles depending on the distance to the impurity (see correlations on Fig.~\ref{fig:Correlations_1imp}) and alternating signs, similarly to Friedel oscillations already evidenced in other continuous spin models with single vacancies\cite{Consoli2024}. 
This phenomena is similar to the selection of the least-collinear order\cite{Maryasin2013, PhysRevLett.119.047204}, also observed on the $J_1J_2$HS model\cite{PhysRevB.105.024201, PhysRevB.104.054201, PhysRevB.86.184432}, explained by an effective positive biquadratic interaction $(\mathbf S_i\cdot\mathbf S_j)^2$ between neighboring spins.
However, with a unique impurity, the $A$ and $C$ quasi-Néel orders remain mutually independent and we still expect them to align collinearly at inifinitesimal temperatures, a partial survival of the initial ObTD. 
With one impurity, we have only two perpendicular spin directions, which is not enough to specify a chirality. 
On the $J_1J_2$HS, the only two independent Néel orders prevent the apparition of a $\mathbb Z_2$ chirality for Heisenberg spins: only a vectorial chirality can be defined, which is a continuous order parameter. However, with $XY$ spins, the chirality is scalar (along $z$) and gives rise to an Ising-like phase transition\cite{Henley1989}. 
In the $J_1-J_3$HK model, with 3 sublattices and with vacancies on all types of sites, the 3 Néel orders tend to orient mutually perpendicularly, allowing to determine a local spin triedra with a scalar chirality. 

We performed extensive Monte Carlo simulations for site dilution $0\leq p\leq0.2$ and linear system size $L$ from $24$ to $128$ (the number of sites is $N=3L^2$). 
To minimize relaxation times, which can be significant at low temperatures, we employed a parallel tempering method combined with a local update provided by the heat bath algorithm \cite{Miyatake1986}. 
Our numerical study focuses on the exchange parameters $J_1=-1$ and $J_3=1$ (we also investigated $J_1 = -1.2$, resulting in a similar phase diagram\cite{Supp}). 
For each $L$, simulations were performed with a minimum average of 10 disorder realizations.

The phase diagram $T-p$ is displayed in Fig.~\ref{fig:PhaseDiag}: the red dot at $p=0$ corresponds to the already known phase transition of the pure model, belonging to the universality class of the 4-state Potts model\cite{Grison2020} (with critical exponents verifying $\alpha/\nu = 1$, $\gamma/\nu = 7/4$, and $1/\nu = 3/2$), as clearly shown by finite-size scaling analysis (FSSA), while the blue dots is the emergent chiral transition with the expected exponents in the Ising universality class $\alpha/\nu = 0$, $\gamma/\nu = 7/4$, and $1/\nu = 1$.
We use the following relations for the maxima of the specific heat $C_V$ and of the (connected) susceptibility $\mathcal S_{X}$ of the order parameter $X$ (with $X=\Sigma$ or $\chi$) to determine the exponents of the phase transitions\cite{Rieger1993}:
\begin{equation}
	\label{eq:FFSA}
\begin{split}
C_V^{\rm max}(L)
&= \frac{\overline{\langle \mathcal H^2\rangle}-\overline{\langle \mathcal H\rangle^2}}{NT^2}\simeq a L^{\alpha/\nu} + b,
\\
\mathcal S_{X}^{\rm max}(L)
&=\frac{\overline{\langle X^2\rangle}-\overline{\langle X\rangle^2}}{NT} \simeq c L^{\gamma/\nu} + d, 
\\
T^{Y^{\rm max}}(L)
&\simeq e L^{-1/\nu} + T_c,
\end{split}
\end{equation}
where brackets denote thermal averages and overlines are averages over disorder realizations.
$T^{Y^{\rm max}}$ is the temperature of the maximum of $Y$, where $Y$ is $C_V$, $\mathcal S_\Sigma$ or $\mathcal S_\chi$, and $T_c$ is the critical temperature.

\begin{figure}[t]
	\includegraphics[width=0.238\textwidth]{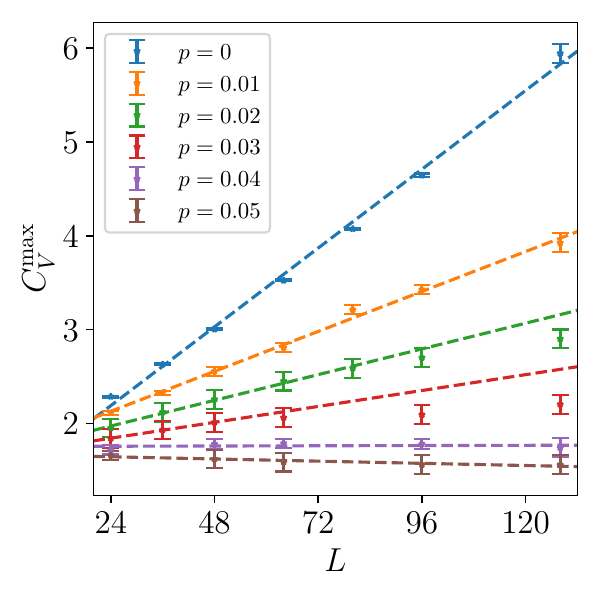}
	\includegraphics[width=0.238\textwidth]{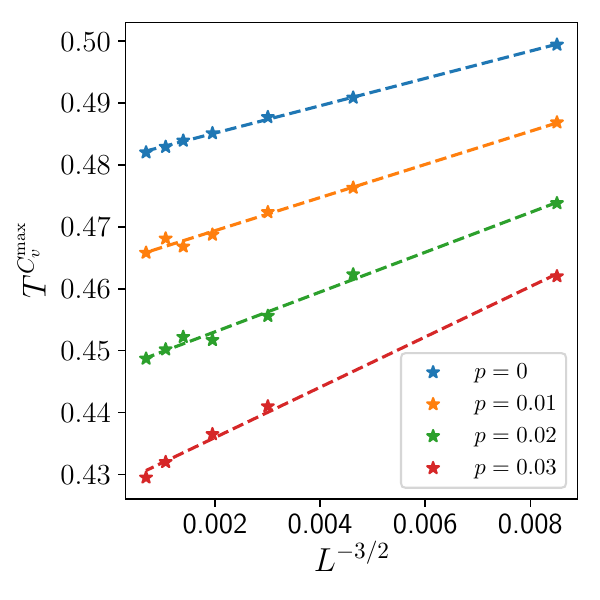}
	\caption{(left) Maxima of heat capacity $C_V^{\rm max}$ versus the linear lattice size $L$ and (right) $T_c^{C_V}(L)$ versus $L^{-3/2}$, for several dilution rates $p$. 
	}
	\label{fig:Scaling}
\end{figure}
\begin{figure}[t]
	\includegraphics[width=0.238\textwidth]{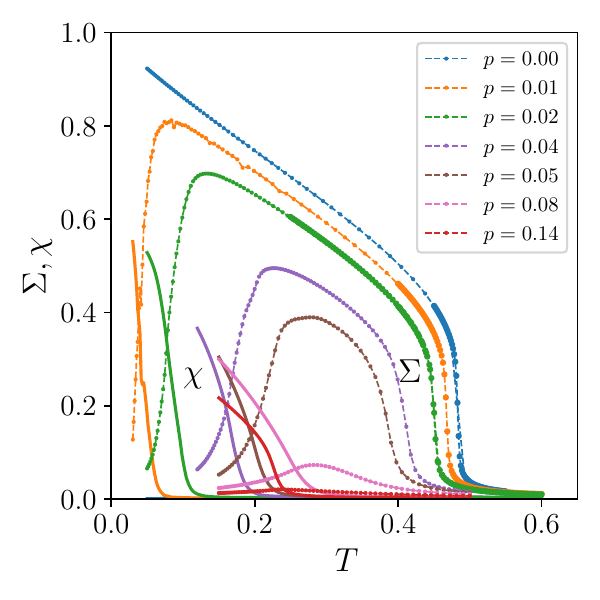}
	\includegraphics[width=0.238\textwidth]{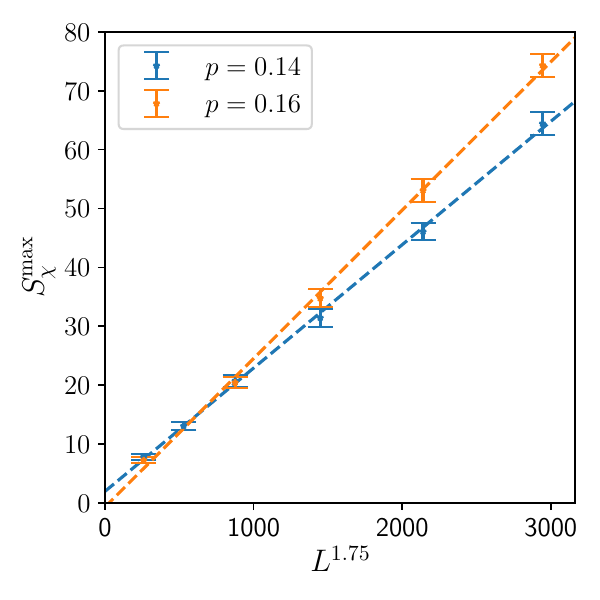}
	\caption{(left) Order parameter $\Sigma$ (increasing and decreasing with $T$) and $\chi$ (immediately decreasing) versus $T$ for different dilutions and $L=96$. (right) Maximum of the susceptibilty $\mathcal S_\chi^{\rm max}$ versus $L^{1.75}$. The dotted lines correspond to linear fits.
	}
	\label{fig:Scaling2}
\end{figure}

We now discuss the phase diagram for $p\neq0$. 
For low impurity rate $p<0.05$, 
we have at first sight all the indications of a phase transition connected to the $K_4$ transition at $p=0$ (green points on Fig.~\ref{fig:PhaseDiag}) in the same universality class as the pure model. $T^{C_V^{\rm max}}(L)$ obeys Eq.~\eqref{eq:FFSA} with a $T_c$ that rapidly decreases with $p$ (Fig.~\ref{fig:Scaling} right)
and we observe a linear dependence of $C_V^{\rm max}(L)$ (Fig.~\ref{fig:Scaling} left).
However, this behavior of $C_V$ is only valid up to a $L^*$ that decreases with $p$ (compare $p=0.02$ and $p=0.03$). 
For $p\geq0.04$, the amplitude $a$ cancels. 
As explained above, site dilution leads to the existence of local random fields that are expected to destroy the order of the pure model at the lower critical dimension $d=2$\cite{PhysRevLett.35.1399}. 
According to the Imry-Ma argument, order is destroyed by the formation of domains, whose size increases exponentially when the impurity rate decreases, as $e^{C/W}$ where $C$ is a constant and $W$ the strength of the random field\cite{Vojta2019}. 
Thus we expect that the disappearence of the $K_4$ phase transition due to dilution can only be observed at large lattice sizes.  
Note that other studies also report signs of transitions in the presence of effective random fields. 
On the $J_1J_2$HS model, a transition similar to the $p=0$ one is \textit{observed} in presence of site-disorder \cite{PhysRevB.86.184432}, but not in Kunwar et al\cite{PhysRevB.98.024206} where the authors do not go below $p=1/8$. 
However, by modifying the nature of the quenched disorder they observe an Imry-Ma contradicting transition\cite{PhysRevB.98.024206} (see footnote 30), arguing that their lattice-symmetry breaking effect is very small\cite{PhysRevB.98.024206}. 
A final example occurs, but in 3d model\cite{Andrade2018}, where an effective gap at $T>0$ is invoked to overcome the Imry-Ma argument. 

At lower temperatures, a phase transition appears, where we observe simultaneously a rapid decrease of $\Sigma$ ($K_4$ order) and a rapid increase of $\chi$ (chiral order) as $T$ decreases (see Fig.~\ref{fig:Scaling2} left).
In the presence of site disorder, non-collinear configurations are energetically favored, overcoming the entropic advantage of the $K_4$ order at very low temperatures. 
The thermodynamic signature of this transition is the appearance of a small secondary maximum in $C_V$ at $T_c \simeq 4p$ for $p<0.05$, which increases weakly with $L$ (e.g., $C_V^{\rm max}(28) = 1.15$, $C_V^{\rm max}(128) = 1.20$). 
This prevents the determination of a reliable scaling exponent with FSSA, that however remains compatible with the scaling of the dilute Ising model\cite{Dotsenko01011983,Ballesteros1997}.
To obtain stable results in this regime, a disorder average over $20$ realizations is required. 
Results are also also limited to $L\leq 80$, due to thermalization difficulties. 

For $0.05 \leq p \leq 0.1$, $\Sigma$ collapses to zero at all temperatures as $L$ increases\cite{Supp}. 
This behavior becomes more pronounced as $p$ increases, illustrating that site dilution effectively acts as a random field for the $K_4$ order and kills the transition. 
At low temperatures, the chiral order is present below a critical temperature that increases with the dilution, but saturates at $p \simeq 0.1$ (see Fig. \ref{fig:PhaseDiag}). 
For $0.1 \leq p \leq 0.2$, the $K_4$ order is clearly absent, and the chiral order weakens as $p$ increases with a progressive decrease of the critical temperature. 
$C_V$ displays a small but noticeable increase with the system size\cite{Supp}. 
Moreover, the maximum of $\mathcal S_\chi$ still diverges with $L$, even though its value decreases due to the  site dilution, which isolates some of the complete triangles where $\chi$ is defined. 
FSSA unveils that this maximum scales as $L^{1.75}$ which is compatible with the Ising universality class (see Fig.~\ref{fig:Scaling2} right).  

For $p=0.2$, the chiral order parameter collapses with $L$, which can be interpreted as the absence of a phase transition due to disorder overcoming some percolation threshold. 
On the kagome lattice, triangles form a honeycomb lattice and have a probability of $(1-p)^3$ to be fully occupied by magnetic sites. 
Equating in a crude approximation this value with the site percolation threshold of the honeycomb lattice (equal to 0.697\cite{Jacobsen_2014}) gives a critical dilution $p_c=0.11$, but this has to be taken with care as further neighbor interactions complicate this view in term of percolation. 

In summary, we have shown on the example of the  $J_1J_3$HK model with magnetic site vacancies that quenched disorder can lead to the emergence of a specific order (here chiral) in $O(3)$-symmetric model, that was forbidden in the clean phase (with collinear $K_4$ order) and previously only observed in $XY$ models\cite{Henley1989,Andrade2018}. 
This phenomenon can be designated as the \textit{order by quench disorder} mechanism that differs from the usual thermal or quantum order-by-disorder. 
Although this expression has been previously employed\cite{PhysRevLett.119.047204}, our proposal is distinct in that it involves the breaking of time-reversal symmetry in the absence of a magnetic field and in contrast to the pure model. 
A similar birth of chirality is expected for other models such as the triangular or honeycomb lattices with stripe orders\cite{Mulder2010, PhysRevB.83.184401}, due to the lattice decoupling in more than two sublattices (similar to what occurs in the $J_1J_3$HK model).
The exploration of the effect of quenched disorder on classical spin liquids\cite{PhysRevB.110.L020402, Yan2024a, PhysRevLett.70.3812, PhysRevB.111.134413, PhysRevResearch.2.043425, PhysRevLett.133.106501, Andrade2018, Consoli2024} represents another direction for further research.

\section{Acknowledgments}

L.M thanks Eric Andrade for invaluable discussions on the effect of disorder, and the IIT Madras for its hospitality that made these discussions possible. 

\bibliography{disorder}
\end{document}


\title{Supplemental Material to Chiral order emergence driven by quenched disorder }

\author{Coraline Letouz\'e}
\affiliation{Sorbonne Universit\'e, CNRS, Laboratoire de Physique Th\'eorique de la Mati\`ere Condens\'ee (LPTMC), F-75005 Paris, France}

\author{Pascal Viot}
\affiliation{Sorbonne Universit\'e, CNRS, Laboratoire de Physique Th\'eorique de la Mati\`ere Condens\'ee (LPTMC), F-75005 Paris, France}

\author{Laura Messio}
\affiliation{Sorbonne Universit\'e, CNRS, Laboratoire de Physique Th\'eorique de la Mati\`ere Condens\'ee (LPTMC), F-75005 Paris, France}

\begin{abstract}
	The definition of the order parameter $\Sigma$ of the $K_4$ order is given. 
	 
	Simulation results for the \(J_1-J_3\) model are presented in detail for different dilutions \(p\), within the phase diagram of the J\(_1\)J\(_3\) HK model.
	For small $p$, a finite-temperature transition is observed between the paramagnetic phase and the 4-state Potts ($K_4$) ordered phase, with the critical temperature decreasing linearly with dilution. At low temperatures, a second transition occurs, where the $K_4$ order disappears and is replaced by chiral order.
	For $p>0.05$, the $K_4$ order vanishes, and at low temperature, the chiral order is associated with a continuous transition. The critical exponents appear to be compatible with an Ising-type transition. For $p>0.2$, the chiral order also disappears.

\end{abstract}
\date{\today}

\maketitle
\tableofcontents

\section{Definition of order parameter for the $K_4$ order: $\Sigma$}

This definition is also the definition used in \cite{Grison2020}, from which several figures are adapted. 

\begin{figure}[t]
	\centering
	\includegraphics[width=\textwidth]{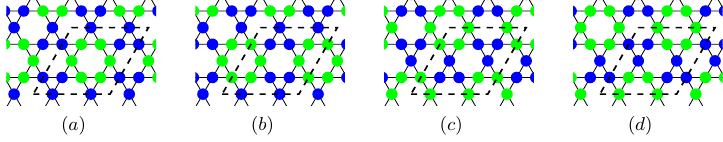}
	\caption{Collinear spin configurations of the 3subAF phase (up to a global $SO(3)$ spin rotation)}
	\label{fig:Coll}
\end{figure}

\begin{figure}[t]
	\centering
	\includegraphics[width=.6\textwidth]{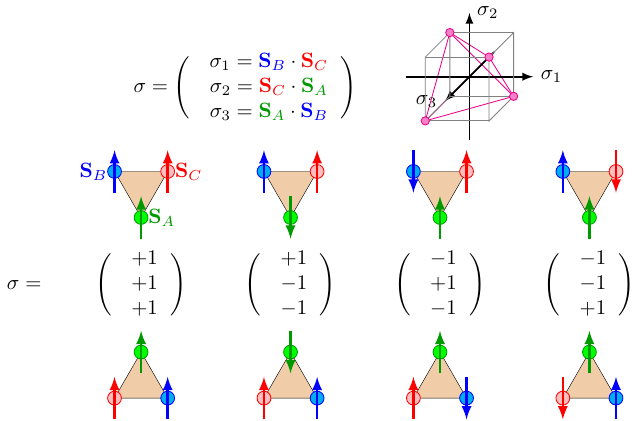}
	\caption{Possible values of $\sigma_{ABC}$ on a triangle with collinear spins (all the possibilities met on Fig.~\ref{fig:Coll} are represented. }
	\label{fig:tetra}
\end{figure}

\begin{figure}[t]
\centering
\includegraphics[width=.6\textwidth]{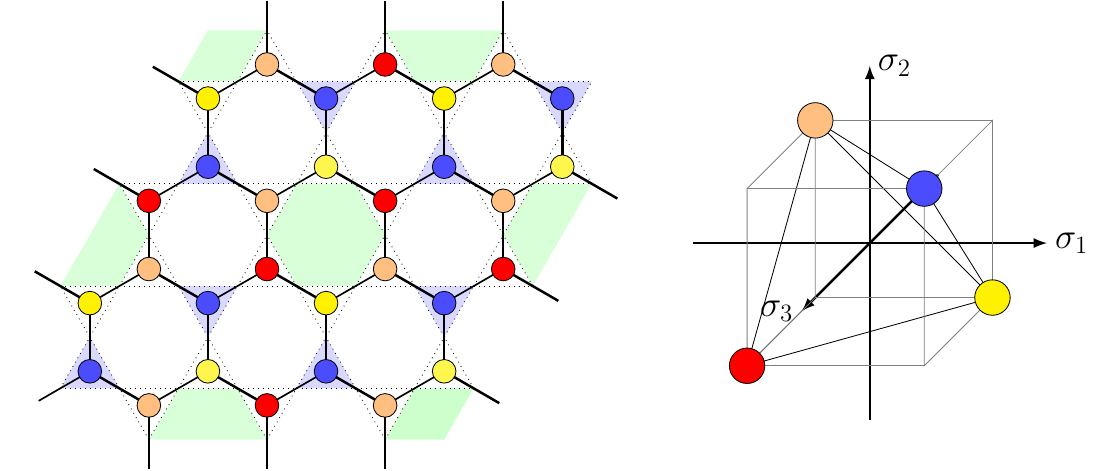}
\caption{Pattern of $\Sigma$ for one of the four configuration of Fig.~\ref{fig:Coll}. The three others are obtained by translation of the lattice. Green hexagons and blue triangles indicates spins with opposite orientations (blue and green arrows on Fig.~\ref{fig:Coll}. }
\label{fig:Hexa_dual}
\end{figure}

As stated in the main article, we define on each triangle (both up and down triangle) a local variable 
\begin{equation}
	\label{eq:orderParameters}
	\bm\sigma_{ABC} 
	= \begin{pmatrix}\mathbf S_B\cdot \mathbf S_C\\ \mathbf S_C\cdot \mathbf S_A\\ \mathbf S_A\cdot \mathbf S_B\end{pmatrix}.
\end{equation}
The four collinear spin orientations of the 3subAF phase (up to a global $SO(3)$ spin rotation) are given on Fig.~\ref{fig:Coll}. 
In these configurations, on any triangle of the kagome lattice, $\bm\sigma_{ABC}$ points towards one of the vertices of a tetrahedron (see Fig.~\ref{fig:tetra}), according to a pattern given in Fig.~\ref{fig:Hexa_dual}
The 4 configurations of Fig.~\ref{fig:Coll} correspond to specific orientations of the tetrahedron, related by the four $K_4$ transformations: the identity and the rotations by $\pi$ around the $\bm\sigma_1$, $\bm\sigma_2$ and $\bm\sigma_3$ axes defined on Fig.~\ref{fig:Hexa_dual}. 
Note that  in any of the collinear configurations, $\bm\sigma_{t_1}$ and $\bm\sigma_{t_2}$ of two fixed triangles $t_1$ and $t_2$ are related by the same $K_4$ transformation, denoted $K_{t_1,t_2}$ such that $\bm\sigma_{t_1}=K_{t_1,t_2}\bm\sigma_{t_2}$, and $K_{t_1,t_2}$ only depends on the color of the triangle in Fig.~\ref{fig:Hexa_dual}.

We define a scalar $\Sigma$ that indicate the strenght of the $K_4$-symmetry breaking. On a finite size lattice, it will be non-zero for a configuration $c$ sampled during the Monte Carlo configuration $\Sigma_c\neq0$, but it will be zero through statistical averages $\langle\Sigma\rangle=0$. 
The two quantities of interest are $\langle|\Sigma|\rangle$ and the susceptibility, directly related to $\langle \Sigma^2\rangle$. A non-zero value of $\langle|\Sigma|\rangle$ in the thermodynamic limit indicates a long-range order, while a divergence in the susceptibility in the thermodynamic limit indicates the occurence of the transition. 

In the 4 configurations, a simple and efficient choice for $\Sigma$ would be to only measure the modulus of the average of $\sigma$ on one triangle over 4 (for example, the yellow triangles in Fig.~\ref{fig:Hexa_dual}). But to have better statistics, we chose another $\Sigma$. 

In the case of the usual Ising antiferromagnetic transition, spins are up on half of the lattice sites, say the $A$ sites, down on the other, the $B$ sites, the order parameter is usually taken as the alternated magnetization $m_A-m_B$ and not only $m_A$. 
To adapt this idea to the $K_4$ phase, we calculate the sum of $\sigma$ over one triangle over 4 for the 4 different colors of triangles of Fig.~\ref{fig:Hexa_dual}. We rotate the four results according to $K_{t_1,t_2}$, where $t_1$ is a reference triangle, and $t_2$ is a representative triangle of the partial sum. 
$\Sigma$ is the the modulus of the sum of these four vectors, and we verify that, once divided by the number of triangle of the lattice, is is 1 in a collinear configuration of Fig.~\ref{fig:Coll}. 


\section{Numerical simulations} 

\begin{figure}[h]
	\includegraphics*[scale=0.6]{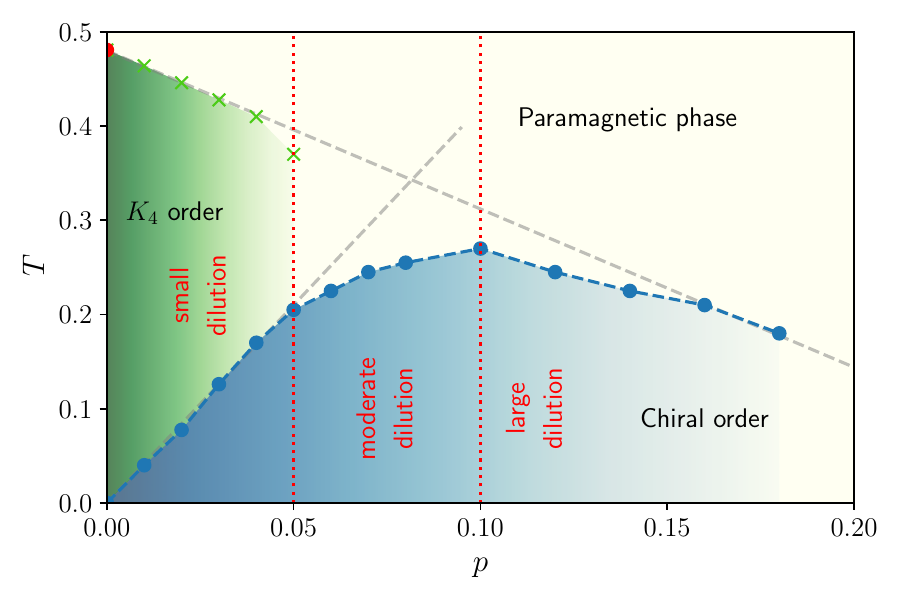}
	\caption{Phase diagram of $J_1J_3KH$ model with $J_3=1$ and $J_1=-1$, presented in the main article.}
	\label{fig:Phasediagram_Regime}
\end{figure}

We distinguish 3 regimes: small, moderate and large dilution, according to the changes in the nature of phase transitions, together with the $p=0$ pure case, and visualized on Fig.~\ref{fig:Phasediagram_Regime} for $J_3=1$ and $J_1=-1$. 

The curves presented for each regime are obtained from several temperature runs performed in parallel, with the parallel tempering method, which regularly attempts exchanges between consecutive temperatures. 
The two order parameters that are measured are defined in the main article and are
\begin{itemize}[noitemsep, nolistsep]
\item $\Sigma$, associated to a broken $K_4$ symmetry (or equivalently $q=4$ Potts order), 
\item and $\chi$, the chirality associated to broken time reversal symmetry and non-coplanarity of the configuration. 
\end{itemize}
For each of order parameter $X$, the connected susceptibility $\mathcal S_X$ is also evaluated using the average of $\Sigma^2$ and $\chi^2$ for each disorder realisation (formula in the main article). 

The linear size $L$ of the lattice is related to the number of kagome sites $N$ via $N=3L^2$. 

\subsection{Pure model} 
\begin{figure}[h!]
	\includegraphics*[scale=0.55]{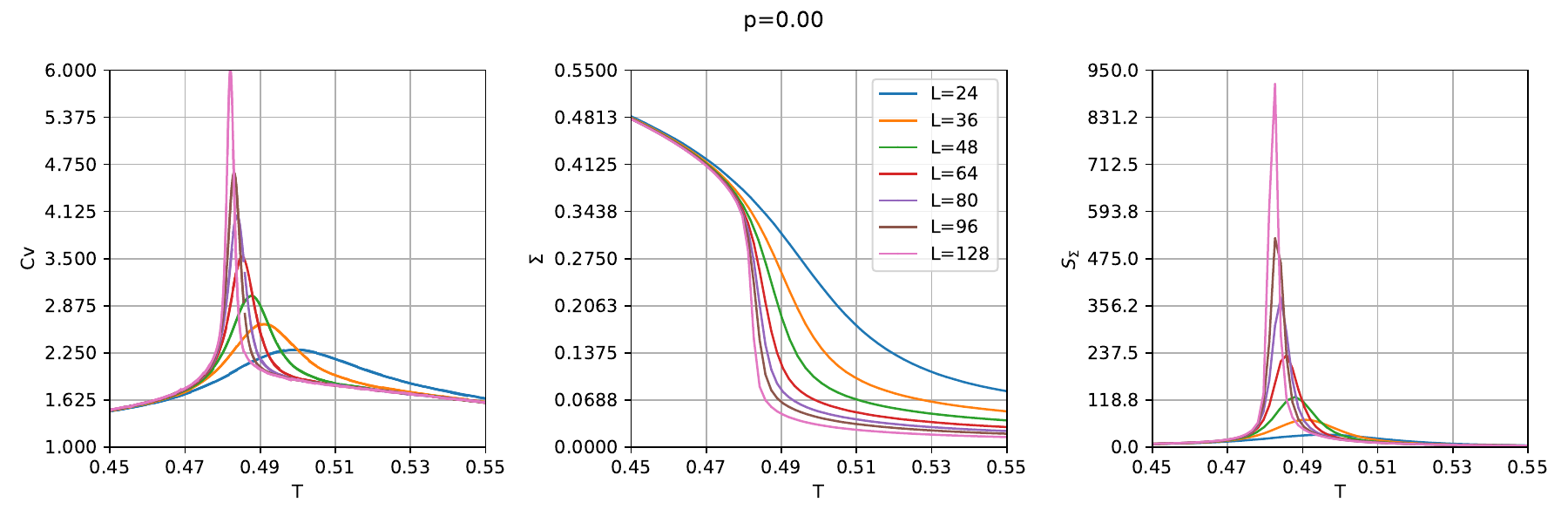}
	\caption{(From left to right) Heat capacity $C_V$, 
		 $K_4$ order parameter $\Sigma$, 
		chiral susceptibility $\mathcal S_\chi$ for the pure model ($p=0$). }
	\label{fig:pure}
\end{figure}

The pure model ($p=0$) exhibits a finite-temperature phase transition already detailed in \cite{Grison2020}, where both the specific heat $C_V$ and the susceptibility $\mathcal S_\Sigma$ diverge with $L$ (see Fig. \(\ref{fig:pure}\)). Finite-size scaling analysis indicates that the critical exponents are consistent with those of the bidimensional four-state Potts model, as expected. Simulations were conducted for different lattice sizes, \(L = 24, \dots, 128\). Figure \(\ref{fig:pure}\) shows the evolution of \(C_V\), \(\Sigma\), and \(\mathcal S_{\Sigma}\).
We also performed simulations at low temperature ($T=0.05$), but the chirality $\chi$ remains always close to zero, which confirms that the order-by-disorder (ObD) mechanism only involves the $\Sigma$ order parameter.

\subsection{Small dilution}

\begin{figure}[h]
	\includegraphics*[scale=0.5]{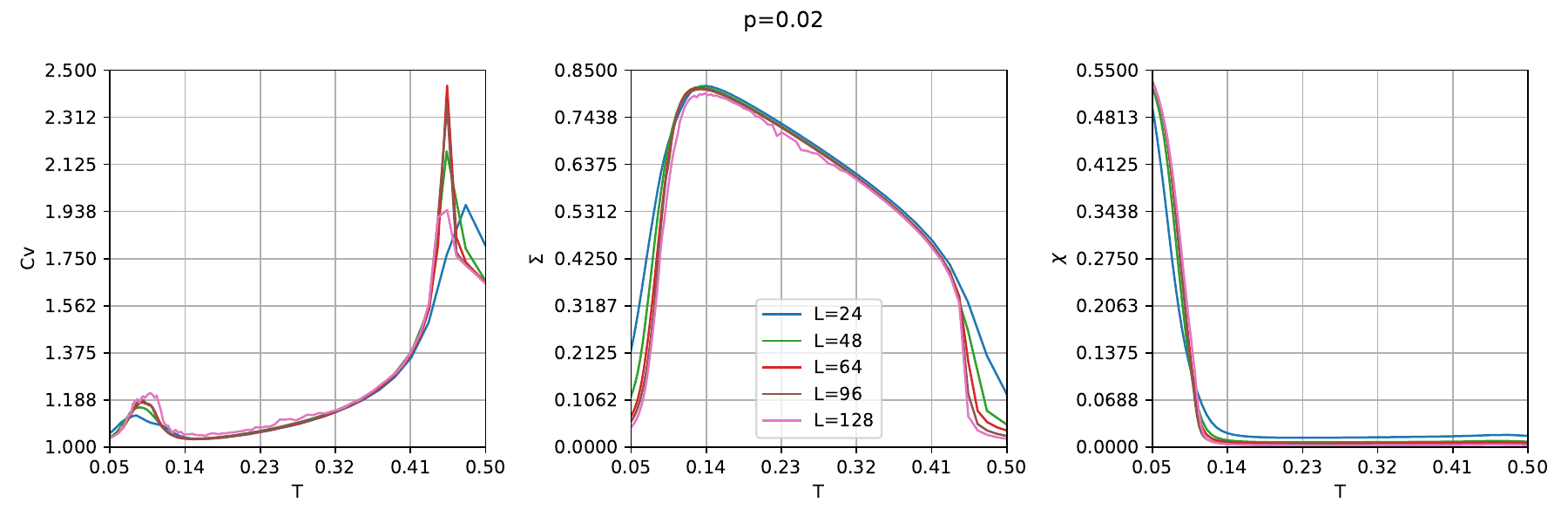}
	\caption{From left to right: heat capacity $C_V$, 
		order parameters $\Sigma$ and $\chi$ for $p=0.02$.}
	\label{fig:smalldilution}
\end{figure}
\begin{figure}
	\includegraphics*[scale=0.5]{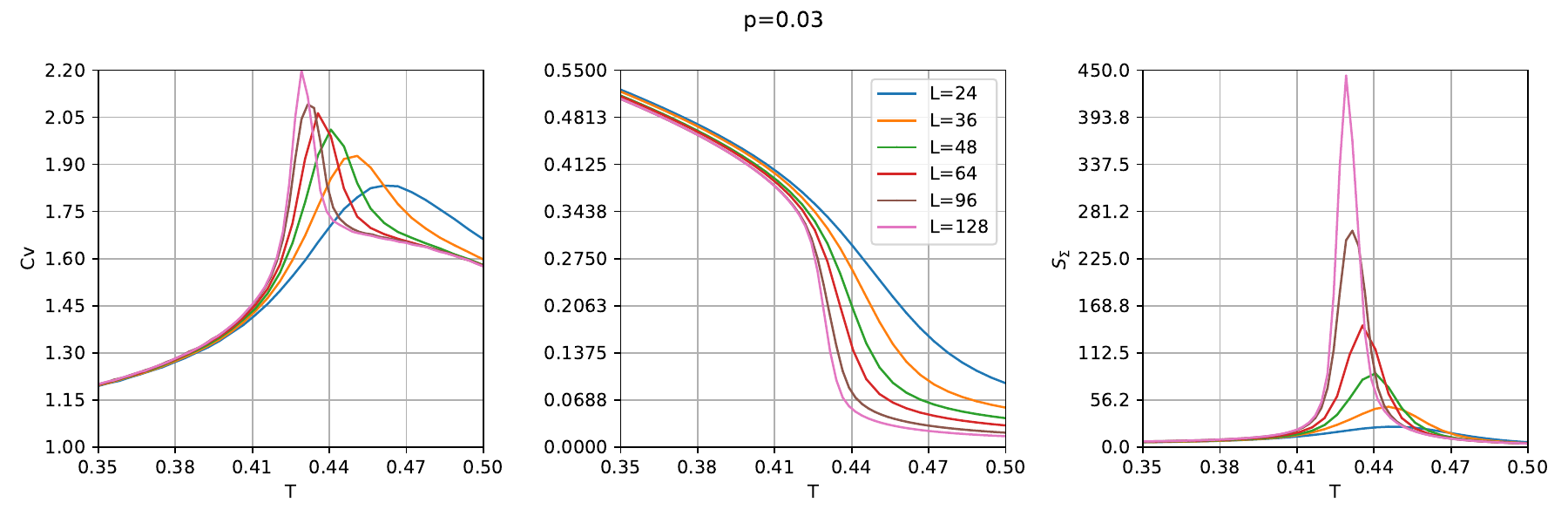} 
	\includegraphics*[scale=0.5]{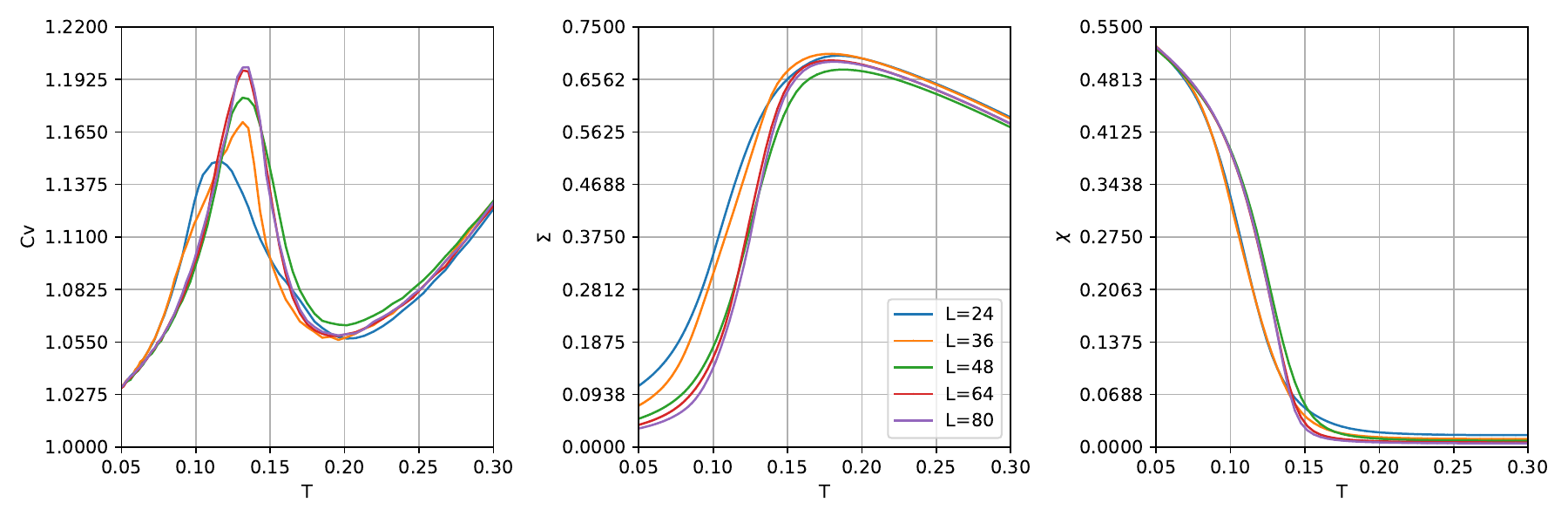} 
	\caption{ Results for $p=0.03$. Top: $C_V$, $\Sigma$ and $\mathcal S_\Sigma$ versus $T$ near the high-temperature peak in $C_V$, bottom: $C_V$, $\Sigma$ and $\chi$ for $T$ near the low-temperature phase transition. }
	\label{fig:smalldilution2}
\end{figure}

We consider the small dilution region ($0 < p < 0.05$). Fig. \ref{fig:smalldilution} and \ref{fig:smalldilution2} present the simulation results respectively for $p=0.02$ and $p=0.03$ and for different system sizes $L$. 

Fig.~\ref{fig:smalldilution} shows $C_V$, $\Sigma$, and $\chi$ versus $T$ for $p=0.02$. 
Two successive event occur for decreasing temperatures in this range of dilutions, corresponding to the two peaks in $C_V$:  
\begin{itemize}[noitemsep, nolistsep]
\item At high temperatures ($T\simeq0.45$), the peak corresponds to the emergence of $K_4$ order (at least at finite size), while the chiral order parameter $\chi$ remains zero. 
Note that the critial temperature is lower than for the pure model ($T_c(p=0)=0.48$). 
\item At lower temperatures ($T\simeq0.08$), a transition occurs, marked by a sudden drop in $\Sigma$ to a small value, accompanied by a rapid increase in $\chi$.  
\end{itemize}

Fig.~\ref{fig:smalldilution2} illustrates again these two successive events for a sightly larger $p=0.03$, but focuses separately on the high and low-temperature regions. 

The high-$T$ plots show $C_V$, $\Sigma$ and the susceptibility $\mathcal S_\Sigma$ as functions of $T$ for different system sizes $L$. 
We explain in the main text why the high temperature peak in $C_V$ is not associated to a phase transition but to a crossover.

The low-$T$ plots show $C_V$, $\Sigma$ and $\chi$ and highlight the low-temperature phase transition. Here, $C_V$ shows little variation with the system size. Near the transition, $\Sigma$ decreases significantly when $T$ decreases, while the chiral order parameter $\chi$ — initially zero — increases rapidly.  

\begin{figure}[h]
	\centering
	\includegraphics*[scale=0.50]{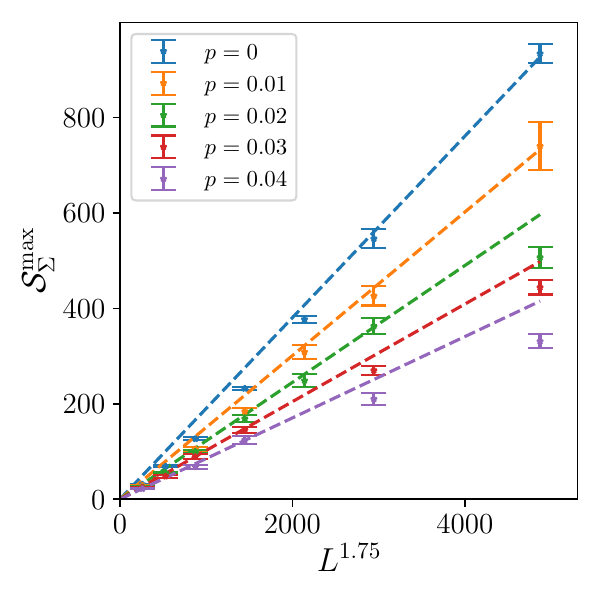}
	\caption{ $\mathcal S_\Sigma^{\rm max}$ versus the linear size $L$ for several dilutions.}
	\label{fig:SusSigma}
\end{figure}

We give the evolution of the maximum of $\mathcal S_\Sigma$ versus $L$ in Fig.~\ref{fig:SusSigma}.

\subsection{Moderate dilution}
\begin{figure}[h]
	\centering
	\includegraphics*[scale=0.60]{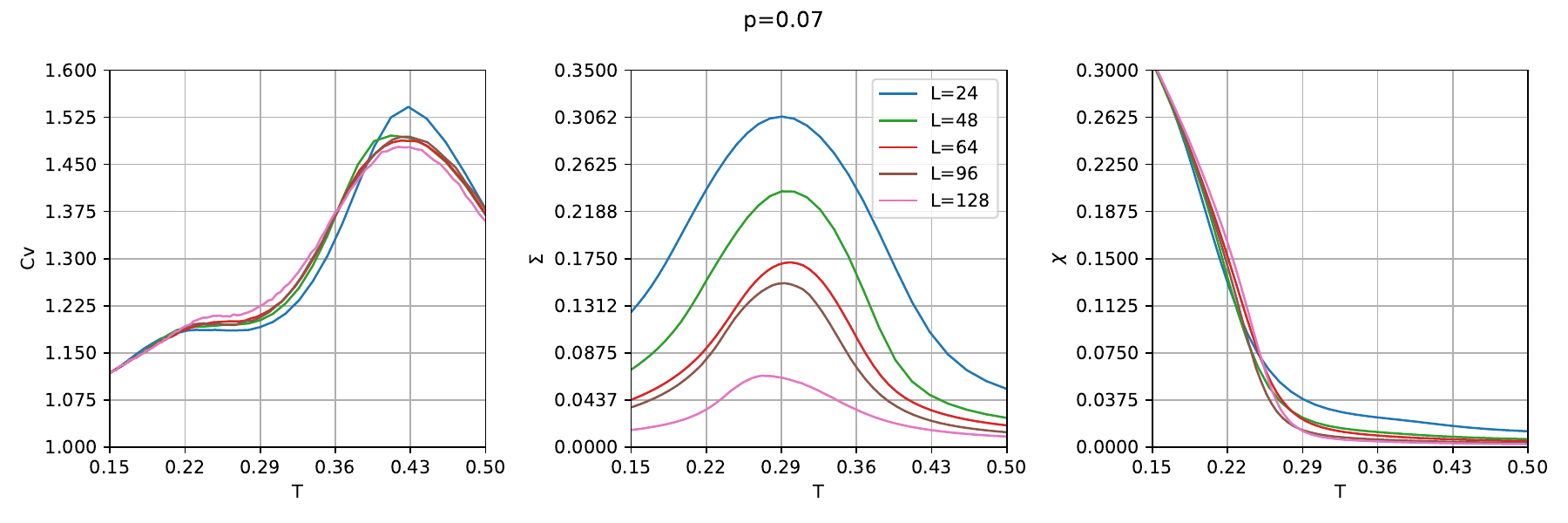}\
	\caption{ (From left to right) $C_V$, $\Sigma$ and $\chi$ for $p=0.07$.  }
	\label{fig:Moderatedilution}
\end{figure}

For \(0.05 \leq p \leq 0.1\), $\Sigma$ tends to zero for any temperature as $L$ increases.
The example of $p=0.07$ is given in  Fig. \(\ref{fig:Moderatedilution}\). 
The low-$T$ peak of $C_V$ disappears, leaving only a shoulder. 
However, the susceptibility $\mathcal S_\chi$ continues to increase with $L$ (not shown).


\subsection{"Large" dilution}
\begin{figure}[ht]
	\centering
	\includegraphics*[scale=0.60]{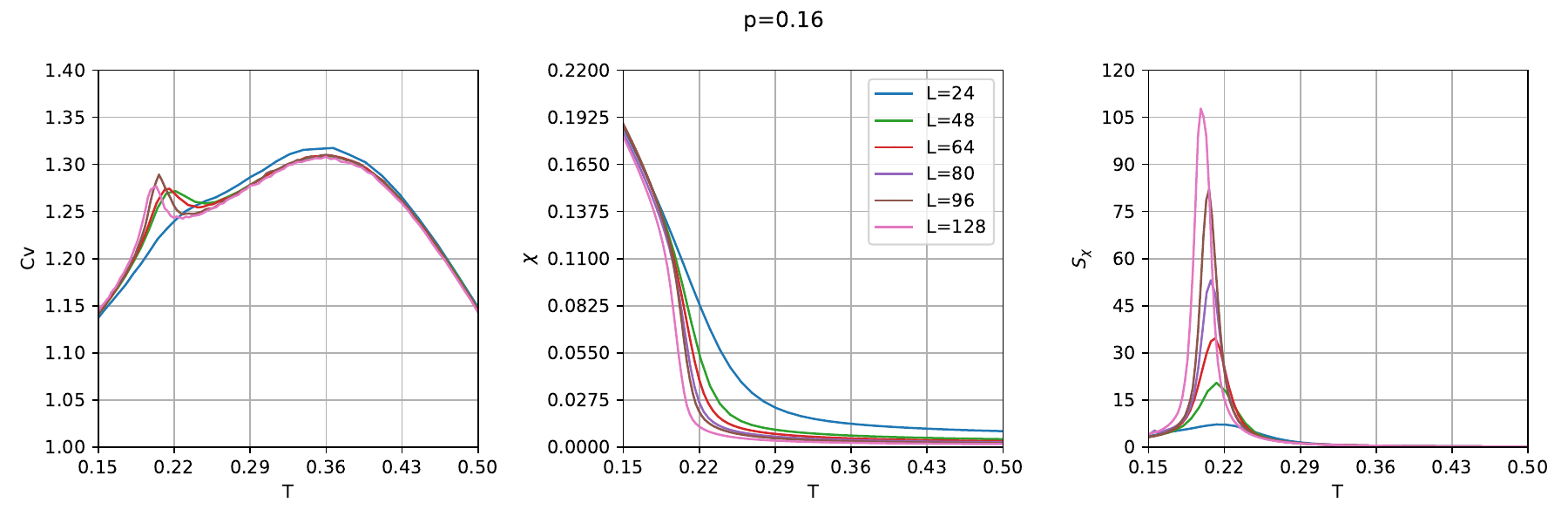}
	\caption{ (From left to right) Heat capacity $C_V$, 
		chiral   order parameter $\chi$ , and the susceptibility $\mathcal S_\chi$ for $p=0.16$.}
	\label{fig:Largedilution}
\end{figure}

For \(0.1 < p < 0.2\), the situation becomes simpler. 
Fig.~\(\ref{fig:Largedilution}\) shows the evolution of \(C_V\), \(\chi\), and the susceptibility $\mathcal S_\chi$ versus $T$ for different lattice sizes $L$. 
The low-$T$ transition is now more clearly identified by a maximum in $C_V$ that increases with $L$, even if the evolution is quite slow.

Moreover, in this region, the chiral transition occurs at a temperature that decreases with dilution. 
Since the chiral order resides on triangles that are free of site dilution, the density of available triangles decreases rapidly with $p$. 
To determine the universality of the transition, we consider  the maximum of the chiral susceptibility $\mathcal S_\chi$, which, as expected, exhibits a power-law behavior with an exponent consistent with an Ising transition (see Fig~.6 of the article).

When $p$ increases,  the intensity of the order parameter diminishes. 
Finally, simulations for \(p \geq 0.2\) lead to a chiral order parameter that decreases with the system size. 
This behavior can be interpreted as related to some percolation property for the triangles with three occupied sites (but the exchanges between first and third neighbors complicate this picture).

\subsection{Phase diagram of the \texorpdfstring{$J_1J_3$}{J1J3}HK model with \texorpdfstring{$J_1=-1.2$}{J1 equal -1.2}}

In all previous results, we have chosen $J_3=1$ and $J_1=-1$. 
In order to enlarge our study, we now consider the effect of disorder when the ferromagnetic coupling is larger $J_1=-1.2$.

\begin{figure}
	\includegraphics*[scale=0.6]{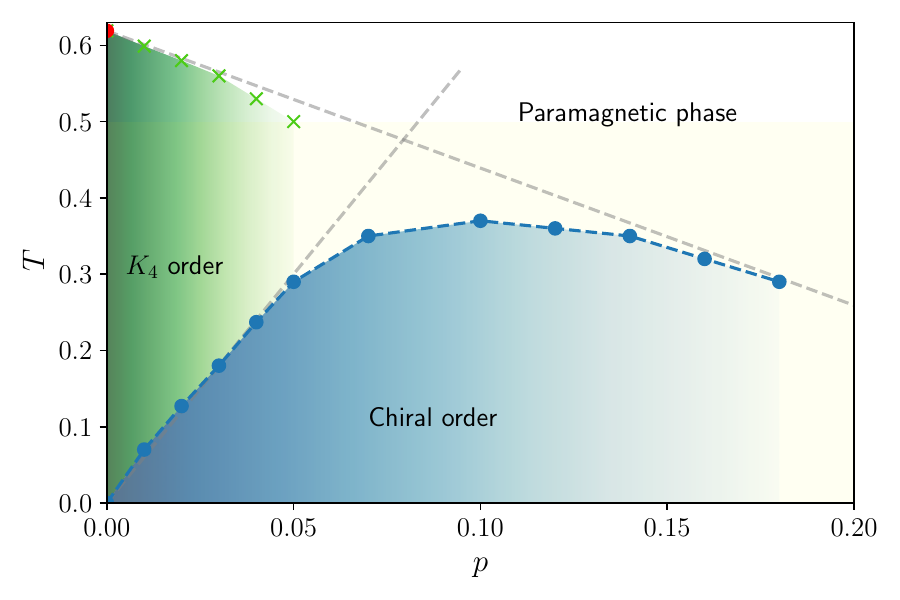}
	\caption{Phase diagram of $J_1J_3KH$ model with $J_3=1$ and $J_1=-1.2$.}
	\label{fig:Phasediagram1.2}
\end{figure}

For the pure model, the critical temperature is shifted to a higher value compared to the pure model with $J_1=-1$. 
By introducing increasing disorder dilution, one obtains a very similar phase diagram, with the same universality properties for each effective phase transition (see Fig. \ref{fig:Phasediagram1.2}).

\section{Environnemental impact of the project}
The numerical simulations were performed on the LPTMC-LKB-LJP cluster, during approximatively 40000d.core, equivalent to 2800kg  eq.CO$_2$, using the approximate rate of 3.6g.core$^{-1}$.h$^{-1}$\cite{berthoud:hal-02549565}.

\bibliography{disorder}